\documentstyle[12pt]{article}
\begin{document}
\hyphenation{me-di-um as-su-ming pri-mi-ti-ve pe-ri-o-di-ci-ty
ma-trix e-le-me-nts fol-lows nu-me-ri-cal co-ef-fi-ci-ents re-du-ced}
\title{
\hfill{\small IPNO/TH 94-20}\\
\hfill{\small April 1994}\vspace*{0.5cm}\\
\sc
%SINGLE-PARTICLE DENSITY OF STATES FOR THE AHARONOV-BOHM POTENTIAL
%AND INSTABILITY OF MATTER WITH ANOMALOUS MAGNETIC MOMENT
Single-particle density of states for the Aharonov-Bohm potential
and the instability of
matter with anomalous magnetic moment
in 2+1 dimensions
\vspace*{0.3cm}}
\author{\sc Alexander Moroz\thanks{e-mail address :
{\tt moroz@ipncls.in2p3.fr}}
\,\thanks{Unit\'{e} de Recherche des Universit\'{e}s Paris XI et Paris VI
associ\'{e}e au CNRS}
\vspace*{0.3cm}}
\date{
\protect\normalsize
\it Division de Physique Th\'{e}orique, IPN\\
\it Univ. Paris-Sud, F-91 406 Orsay Cedex\\
\it France
}
\maketitle
\begin{center}
{\large\sc abstract}
\end{center}
In the nonrelativistic case I found that whenever the  relation
$mc^2/e^2 <X(\alpha,g_m)$ is satisfied, where $\alpha$ is a flux,
$g_m$ is magnetic moment,
and $X(\alpha,g_m)$ is some function that is nonzero only for
$g_m>2$ (note that $g_m=2.00232$ for the electron),
then the matter is unstable against formation of the flux $\alpha$.
The result persists down to $g_m=2$ provided the Aharonov-Bohm
potential is supplemented with a short range attractive potential.
I also show that whenever a bound state is present in the spectrum it
is always accompanied by a resonance with the energy proportional to
the absolute value of the binding energy.
In the relativistic case one again finds the resonance when
the bound state is present but the instability
disappear as long as the minimal coupling is
considered. For the Klein-Gordon equation with the Pauli coupling
which exists in (2+1) dimensions without any reference to a spin
the matter is again unstable for $g_m>2$.
The results are obtained by calculating the change
of the density of states induced by the Aharonov-Bohm potential.
The Krein-Friedel formula
for this long-ranged potential is shown to be valid
when supplemented with zeta function regularization.

\vspace*{0.3cm}

%\vspace{0.6cm}

\thispagestyle{empty}
\baselineskip 20pt
\newpage
\setcounter{page}{1}
%%%%%%%%%%%%%%%%%%%%%%%%%%%%%%%%%%%%%%%%%%%%%%%%%%%%%%%%%%%%%%%%%%%%%%%%
%%%%%%%%%%%%%%%%%%%%%%%%%%%%%%%%%              %%%%%%%%%%%%%%%%%%%%%%%%%
\noindent
In this letter a change of the density of states in the whole space
is calculated
for the Schr\"{o}dinger, the Klein-Gordon, and the Dirac equations
with the Aharonov-Bohm (AB) potential
$A_r=0$, $A_\varphi=\Phi/2\pi r$. This enables to discuss
the stability properties of matter against the spontaneous creation
of a magnetic field. In the relativistic case
and for the normal magnetic moment
we reconfirm previous result, known as the {\em diamagnetic inequality}
\cite{HSS} that the matter is {\em stable}.
The latter was proven under the assumption of minimal
coupling which implicitely assumes the normal magnetic moment.
However,  in the nonrelativistic case
a window exists for the magnetic moment $g_m>2$
in which the inequality is {\em violated} leading to the
{\em instability} of matter against a magnetic field formation.
The reason is the formation of bound states which decouple from
the Hilbert space by taking away negative energy.

Note that one has the unitary equivalence between a spin $1/2$ charged
particle in
a 2D magnetic field and a spin $1/2$ neutral particle with an anomalous
magnetic moment in a 2D electric field \cite{OO}.
One also has a  formal similarity between the scattering of electron
in the AB potential and in the spacetime of a gravitational vortex
in $2+1$ dimensions \cite{GJ}.
Moreover, one encounters
the AB potential (of non-magnetic origin) in the cosmic-string
scenarious \cite{AW} and our results apply
to the this cases as well.

{\bf 1.} {\em The Schr\"{o}dinger equation}.-
By using the separation of variables, assuming $e=-|e|$,
the total Hamiltonian is written as a direct sum of channel
radial Hamiltonians $H_l$ \cite{AB,R}
\begin{equation}
H_l=-\frac{d^2}{dr^2}-\frac{1}{r}\frac{d}{dr}+\frac{\nu^2}{r^2},
\label{schrham}
\end{equation}
where $\nu=|l+\alpha|$, $\alpha$ being
the total flux $\Phi$ in the units
of the flux quantum $\Phi_o=hc/|e|$, with the spectrum given by
\begin{equation}
\psi_l(r,\phi)=J_{|l+\alpha|}(kr)e^{i\phi},
\label{regpsi}
\end{equation}
with $k=\sqrt{2mE/\hbar^2}$.
The eigenfunction expansion for the Green function
in the polar coordinates ${\bf x}=(r_x,\phi_x)$
is
\begin{equation}
G({\bf x},{\bf y},E)=\frac{m}{\pi\hbar^2}\int_o^\infty
\frac{k dk}{q^2-k^2}\sum_{l=-\infty}^\infty
e^{il(\phi_x-\phi_y)}
J_{|l+\alpha|}(kr_x)J_{|l+\alpha|}(kr_y).
\label{gr:def}
\end{equation}
The normalization is chosen such that
$-(1/\pi)\mbox{ImTr}\,G({\bf x},{\bf x},E+i\epsilon)$ gives
the two dimensional
free density of states $\rho_o(E)=(mV/2\pi\hbar^2)dE$ for $\alpha\in Z$.
In this case the sum in (\ref{gr:def}) can be taken exactly
by means of Graf's addition theorem (\cite{AS}, relation 9.1.79).
By taking the integral (assuming $q^2=q^2+i\epsilon$, and $r_x<r_y$)
one finds
\begin{equation}
G({\bf x},{\bf y},E)=\frac{m}{\pi\hbar^2}
\int_o^\infty \frac{k dk}{q^2-k^2} J_o(k|{\bf x}-{\bf y}|) =
-i\frac{m}{2\hbar^2}H^1_o(q|{\bf x}-{\bf y}|).
\label{gr:res}
\end{equation}
One can also take  the integral
\begin{equation}
\int_o^\infty \frac{k dk}{q^2+i\epsilon-k^2} J_\nu(kr_x)
J_\nu(kr_y)=-\frac{\pi i}{2}J_\nu(qr_x)H^1_\nu(qr_y)
\label{int}
\end{equation}
at first (which is valid for nonintegral $\nu$ too) and then to take
the remaining sum by Graf's theorem with the same result (\ref{gr:res}).
{}From the `formal scattering' point of view taking the residuum
at $k=|q|+i\epsilon$ corresponds to choosing the outgoing
boundary conditions.
{}From the point of view of $L^2(R^2)$ it corresponds
to taking the boundary value of the resolvent operator on
the upper side of the cut at $[0,\infty)$ in the complex energy plane
\cite{RR}. The limiting value of
the resolvent operator on the lower side of the cut is
the complex conjugate of (\ref{gr:res}),
the discontinuity across the cut
%\begin{equation}
$G({\bf x},{\bf x},E_+)-G({\bf x},{\bf x},E_-)=-i(m/\hbar^2)$,
%\end{equation}
and
\begin{equation}
-(1/\pi)\,\mbox{Im}\, G({\bf x},{\bf x},E_+)= m/2\pi\hbar^2
\end{equation}
which confirms our normalization.

Whenever $\alpha\not\in Z$  Graf's theorem  cannot be used.
To proceed further with  this case we use the fact that
(\ref{int}) has an analytic continuation on the imaginary axis
in the complex {\em momentum} plane
\begin{equation}
\int_o^\infty \frac{k dk}{q^2+k^2} J_\nu(kr_x)
J_\nu(kr_y)=\frac{\pi i}{2}J_\nu(iqr_x)H^1_\nu(iqr_y)
=I_\nu(qr_x)K_\nu(qr_y),
\end{equation}
where $I_\nu$ and $K_\nu$ are modified Bessel functions.
To sum over $l$ one uses the integral representation
of these functions and following the steps of \cite{MRS} one
can separate the $\alpha$-dependent contribution
\begin{equation}
G_\alpha({\bf x},{\bf x},M)-G_o({\bf x},{\bf x},M)=
\frac{m}{\hbar^2}
\frac{\sin(\eta\pi)}{(2\pi)^2}
\int_{-\infty}^\infty d\vartheta\,\int_{-\infty}^\infty d\omega
e^{-M
r_x(\cosh\vartheta +\cosh\omega)}\,\frac{e^{\eta(\vartheta-\omega)}}
{1+e^{\vartheta-\omega}},
\end{equation}
where $M=-i\sqrt{2mE/\hbar^2}$ and $\eta$ is the nonintegral part of
 $\alpha$, $0\leq \eta<1$.
After taking the trace over spatial coordinates,  using formulae
3.512.1 and 8.334.3 of \cite{GR}, and returning back
on the real momentum axis  one finally finds
\begin{equation}
\mbox{Tr}\,[G_\alpha({\bf x},{\bf x},E)-G_o({\bf x},{\bf x},E)]=
-\frac{1}{2}\,\eta(1-\eta)\,\frac{1}{E}
\end{equation}
which gives the change of the density of states in the {\em whole space}
\begin{equation}
\rho_\alpha(E)-\rho_o(E)=- \frac{1}{2}\,\eta(1-\eta)\,\delta(E)
\label{denreg}
\end{equation}
that confirms \cite{CGO}, where it was obtained in the context of anyonic
physics.

The channel Hamiltonians $H_l$  for which $|l+\alpha|<1$
admit a one-parametric family of self-adjoint extensions \cite{R,RR}.
We shall consider the situation with
bound states
\begin{equation}
\psi_l(r,\phi)=K_{|l+\alpha|}(\kappa_l r)e^{i\phi}
\end{equation}
of energy $E_l=-(\hbar^2/2m)\kappa_l^2$
at $l=-n$ and $l=-n-1$ channels,
with $n=[\alpha]$ denoting the nearest  integer {\em smaller} than
$\alpha$.
In the presence of the bound states the scattering states (\ref{regpsi})
have to be modified
\begin{equation}
\psi_l(r,\phi)\rightarrow\psi_l(r,\phi)=(J_{|l+\alpha|}(kr)-A_lJ_{-|l+\alpha|}
(kr))e^{il\phi}
\label{psising}
\end{equation}
This is because $H_l$ has to be necessary  a symmetric operator
what already determines $A_l$ to be $A_l=(k/\kappa_l)^{2\nu}$, i.e.,
{\em energy dependent}.
To calculate the change of the  {\em integrated}
density of states in the whole space
we make use of the Krein-Friedel formula \cite{F}
\begin{equation}
N_\alpha(E)-N_o(E)=-\frac{i}{2\pi}\ln\det\mbox{S},
\label{krein}
\end{equation}
S being the total on-shell S-matrix.
Despite the AB potential is long-ranged we have found that
the Krein-Friedel formula
when combined with $\zeta$-function regularization
can  still be used.
The radial part of the general solution (\ref{psising}) behaves as
\begin{equation}
R_l(r)\sim \mbox{const}\left(e^{-ikr}+\frac{1-A_le^{i\pi|l+\alpha|}}
{1-A_le^{-i\pi|l+\alpha|}}e^{-i\pi(|l+\alpha|+1/2)}
e^{ikr}\right)\hspace*{1cm}(r\rightarrow\infty),
\end{equation}
which determines the $l$-th channel S${}_l$-matrix to be
S${}_l=e^{2i\delta_l}$ with
\begin{equation}
\delta_l=\frac{1}{2}\pi(|l|-|l+\alpha|)+\arctan\left(\frac{\sin(|l+
\alpha|\pi)}{\cos(|l+\alpha|\pi) -A_l^{-1}}\right).
\label{shift}
\end{equation}
Without the presence of bound states ($A_l=0$)
\begin{eqnarray}
\lefteqn{\ln\det\mbox{S}=\sum_{l=-\infty}^\infty 2i\delta_l}\nonumber\\
&&= i\pi\sum_{l=-\infty}^\infty (|l|-|l+\alpha|)=
%\nonumber  \\
i\pi\left.\left[ 2\sum_{l=1}^\infty l^{-s}-\sum_{l=0}^\infty (l+\eta)^{-s}
%\hspace*{1cm}
-\sum_{l=1}^\infty (l-\eta)^{-s}\right]\right|_{s=-1}
\nonumber  \\
&&
 =i\pi\left.\left[2\zeta_R(s)-\zeta_H(s,\eta)-\zeta_H(s,1-\eta)\right]
\right|_{s=-1}=
-i\pi\eta(1-\eta),
\end{eqnarray}
where $\zeta_R$ and $\zeta_H$ are the Riemann and the Hurwitz
$\zeta$-function.
The use of the Krein-Friedel formula (\ref{krein}) then
independently confirms (\ref{denreg}).
Under the presence of bound states  the contribution
of scattering states to $N_\alpha$ for $E\geq 0$ is
\begin{eqnarray}
N_\alpha(E)-N_o(E)= &-&\frac{1}{2}\eta(1-\eta)+\frac{1}{\pi}\arctan
\left(\frac{\sin(\eta\pi)}{\cos(\eta\pi)-(|E_{-n}|/E)^{\eta}}
\right)
\nonumber\\
&-&\frac{1}{\pi}\arctan
\left(\frac{\sin(\eta\pi)}{\cos(\eta\pi)+
(|E_{-n-1}|/E)^{(1-\eta)}}\right),
\label{intsing}
\end{eqnarray}
where $E_{-n}$ and $E_{-n-1}$ are the binding energies
in $l=-n$ and $l=-n-1$ channels. Note that for
$0<\eta<1/2$ we have a {\em resonance}
at
\begin{equation}
E=\frac{|E_{-n}|}{(\cos(\eta\pi))^{1/\eta}}>0.
\end{equation}
The phase shift
$\delta_{-n}(E)$ (\ref{shift}) changes by $\pi$ in the direction
of increasing energy and  the integrated density of states
(\ref{intsing}) has a sharp
increase by one. For $1/2<\eta<1$ the resonance is shifted to the
$l=-n-1$ channel.
$\eta=1/2$ is a special point since resonances are in both channels
at infinity. Since we always are in front of the
resonances the contribution
of {\em arctan} terms in (\ref{intsing})
does not vanish as $E\rightarrow\infty$ but instead gives $-1$.

Different self-adjoint extensions correspond to different
physics inside the flux tube \cite{PG}.
To identify the physics which underlines them
we have considered the situation when the AB potential
is regularized by a {\em uniform} magnetic $B$ field within the radius $R$.
In order for a  bound state to exist the matching equation for
the exterior and interior solutions in the $l$-th channel,
\begin{equation}
x\frac{K_{|l+\alpha|}'(x)}{K_{|l+\alpha|}(x)}=-\alpha+|l|+\alpha
\frac{|l|+l+1+(x^2/2\alpha)}
{|l|+1}\frac{{}_1F_1\left(\frac{|l|+l+3}{2}+(x^2/4\alpha),|l|+2,\alpha\right)}
{{}_1F_1\left(\frac{|l|+l+1}{2}+(x^2/4\alpha),|l|+1,\alpha\right)},
\label{matching}
\end{equation}
with ${}_1F_1(a,b,c)$ the Kummer hypergeometric function \cite{AS},
has to have a solution $x_l=\kappa_l R\neq 0$. However, since the l.h.s.
decreases from $-|l+\alpha|$ to $-\infty$ as $x\rightarrow\infty$ and
the r.h.s. is always positive one finds that
it is impossible unless it is an
{\em attractive} potential $V(r)$ inside the flux tube,
\begin{equation}
V(r)|_{r\leq R}=-\frac{\hbar^2}{2m^2}\frac{\alpha}{R^2}c(R),
\label{tubpot}
\end{equation}
with $c(R)=2(1+\epsilon(R))$, $\epsilon(R)>0$, and $\epsilon(R)\rightarrow 0$
as $R\rightarrow 0$ \cite{BV}.
This amounts to changing $x^2/2\alpha$ by $ x^2/2\alpha -c/2$ on the r.h.s.
of (\ref{matching}). The attractive potential can be either put in
by hands or, when the Pauli equation is considered, as arising
in the magnetic momentum coupling of electrons
with spin  opposite to the direction of magnetic field $B$.
The case of gyromagnetic ratio $g_m=2$ ($\epsilon(R)\equiv 0$)
is a critical point at which (\ref{matching}) has a
solution at $x=0$
for $l=-n$. It is known that there are $]\alpha[-1$  {\em zero modes}
in this case, $]\alpha[$ the nearest integer {\em larger} than $\alpha$
\cite{AC}. The result does  only depend on the total flux $\alpha$ and
 not on a particular distribution of a magnetic field $B$.

Whenever $g_m>2$ (or $g_m=2$ with an attractive potential
$V(r)=-\epsilon/R^2$, $\epsilon>0$ arbitrary small)
the bound states may occur in the spectrum.
They correspond to solutions $x_l>0$ of (\ref{matching}).
In contrast to the zero modes their number {\em does depend}
on a particular distribution of the magnetic field $B$ but
it is less than or equals to  \cite{BV1}
\begin{equation}
\#_b =1 +n +2 [\alpha(g_m-2)/4]
\end{equation}
with $[.]$ as above.
The bound is saturated when one uses  the cylindrical shell regularization
of the AB potential \cite{BV,Hag}. Note that in this case the energy
$E_B$ of magnetic field is {\em infinite} for any $R\neq 0$ in contrast
to homogeneous field regularization when
\begin{equation}
E_B=\frac{1}{2}\pi B^2 R^2= \frac{\Phi^2}{2\pi R^2}\cdot
\label{mgen}
\end{equation}

{\bf 2.} {\em Energy calculations}.-
Up to $g_m=2$ no bound state is present in the spectrum
and the change of the density of the
scattering states is still given by (\ref{denreg}). Zero modes
which may occur for $g_m=2$ are regular at the origin and do not
change phase shifts except for $l=-n$ and $l=-n-1$ channels where
they cause phase shift flip ($A_l^{-1}=0$ in (\ref{shift})) (cf. \cite{Hag})
but in sum they cancel.
Therefore the energy of magnetic field (\ref{mgen})
tends to infinity as $R\rightarrow 0$ (when the total flux $\Phi$ is kept
fixed) the matter is {\em stable} with regard to a spontaneous
creation of the AB field.

When $g_m>2$ then bound states may occur in the spectrum.
Their energy is
\begin{equation}
E_l=-\frac{\hbar^2}{2m}\frac{x_l^2}{R^2},
\label{bsen}
\end{equation}
and tends to $-\infty$ as $1/R^2$ when $R\rightarrow 0$,
 in the same way as the magnetic field energy (\ref{mgen}) goes to $\infty$.
The bound states decouple in the $R\rightarrow 0$ limit from the Hilbert
space $L^2(R^2)$ ($A_l\rightarrow 0$ in (\ref{psising}) in the limit)
and take away the {\em nonperiodicity} of the spectrum
with regard to $\alpha\rightarrow\alpha\pm1$ which
persists for any finite $R$.
What is left behind is nothing but the
{\em conventional} AB problem with the change of the density
of states (\ref{denreg}).  We recall that to have a {\em finite energy} bound
state in the spectrum it has to be the attractive potential (\ref{tubpot})
inside the flux tube.

Bound states solutions $x_l$ for the homogeneous
field regularization
determine the function $X(\alpha,g_m)$,
\begin{equation}
X(\alpha, g_m)=\frac{1}{4\pi\alpha^2}\sum_l x^2_l(\alpha,g_m)\geq 0.
\end{equation}
By comparing the coefficients in front of $1/R^2$ in (\ref{mgen}) and
(\ref{bsen}) one finds that whenever
\begin{equation}
mc^2/e^2<X(\alpha,g_m)
\label{instab}
\end{equation}
the total energy of field and matter altogether goes to $-\infty$
as $R\rightarrow 0$.

{\bf 3.} {\em The Dirac  and the Klein-Gordon equations}.-
A connection with the Schr\"{o}dinger equation is established by
noticing that for the Dirac spinor to belong to
the spectrum its up/down radial component
has to be an eigenfunction with an eigenvalue $k^2$ of
\begin{equation}
H_l=-\frac{d^2}{dr^2}-\frac{1}{r}\frac{d}{dr}+\frac{\nu^2}{r^2}\pm g_m
\frac{\alpha}{r}\delta(r),
\label{dirham}
\end{equation}
with  $g_m=2$ and ($\nu=l+\alpha$/$\nu=l+1+\alpha$) for the (up/down)
component \cite{Hag}. The dispersion relation for $k$ is now changed to
$k=(1/\hbar c)\sqrt{E^2-m^2c^4}$.
The two-component solutions of the massive Dirac equation have
only {\em one degree of freedom} which is reflected in the equality
of up and down phase shifts \cite{GJ}. For the conventional AB problem
$\delta^{u}_l=\delta^{d}_l=\pi\alpha/2$  if $l= -n-1$, and
$\delta^{u}_l=\delta^{d}_l= \pi(|l|-|l+\alpha|)/2$
otherwise.
Using the results of \cite{PG,Th} one finds that for a fixed real $k\neq 0$
the spectrum is {\em symmetric} with respect
to the origin, in accord with  supersymmetry \cite{Th}.
Thus the contribution of scattering states to the density of states
is given by
\begin{equation}
\rho_{\alpha}({\cal E})-\rho_o({\cal E}) = -
\frac{1}{2}\eta(1-\eta)\,\delta({\cal E}^2-m^2c^4).
\label{dirden}
\end{equation}
 The only asymmetry can occur for the threshold \cite{Th} and bound states.
Depending on the orientation of $B$ the threshold states occur
either {\em only} at the upper or {\em only} at the lower threshold \cite{Th}.
Their number equals to the number $]\alpha[-1$ of zero modes of the
Schr\"{o}dinger equation \cite{AC,Th}.

    By considering different self-adjoint extensions one finds that,
in contrast to the Schr\"{o}dinger equation, bound states can only
occur in a {\em single}
channel $l=-n-1$  ($l=-n$ for positive charge)
\cite{PG}.  In the presence of the bound state of
energy $E$ in the $l=-(n+1)$-th channel,
\begin{equation}
\Psi_{E;-n-1}(t,r,\phi)=
\frac{1}{N}
\left(
\begin{array}{c}
\sqrt{mc^2+E}K_{1-\eta}(\kappa r)\\
i\sqrt{mc^2-E}K_{\eta}(\kappa r) e^{i\phi}
\end{array}\right)
e^{-i(n+1)\phi}e^{-i{\cal E}t/\hbar},
\end{equation}
where $\kappa=(1/\hbar c)\sqrt{m^2c^4-E^2}$,
the scattering state with energy ${\cal E}>mc^2$ is
\begin{equation}
\Psi_{{\cal E};-n-1}(t,r,\phi)=
\left(
\begin{array}{c}
\chi^1(r)\\
\chi^2(r)e^{i\phi}
\end{array}\right)
e^{-i(n+1)\phi}e^{-i{\cal E}t/\hbar},
\end{equation}
\begin{equation}
\chi_{-n-1}(r)=\frac{1}{N}
\left(
\begin{array}{c}
\sqrt{{\cal E}+mc^2}[\sin\mu\,J_{1-\eta}(kr)+(-1)^{n+1}
\cos\mu\,J_{-(1-\eta)}(kr)]\\
i\sqrt{{\cal E}-mc^2}[-\sin\mu\,J_{-\eta}(kr)+(-1)^{n+1}
\cos\mu\,J_{\eta}(kr)]
\end{array}\right)
\label{scatdir}
\end{equation}
with  $N$ a normalization factor \cite{PG}.
Bound state energy $E$ again parametrizes different self-adjoint extensions
and,  for given ${\cal E}$, determines the parameter $\mu$
\begin{equation}
\tan\mu =
(-1)^{n+1}\frac{mc^2-E}{(m^2c^4-E^2)^{\eta}}
\frac{({\cal E}^2-m^2c^4)^{\eta}}{{\cal E}-mc^2}\cdot
\label{tan}
\end{equation}
By comparing (\ref{scatdir}) with (\ref{psising}) one finds
a change of the conventional phase shifts
\begin{equation}
\delta =-\arctan\frac{\sin(\eta\pi)}{\cos(\eta\pi)+
(-1)^n\tan\mu}\cdot
\end{equation}
When the threshold state ${\cal E}=mc^2$ is present $\tan\mu=0$ and
one has a phase shift flip, in accord with \cite{Hag}.
%For $\eta>1/2$ ($\eta<1/2$) $\delta^u$ ($\delta^d$)
%tends to zero as ${\cal E}\rightarrow\infty$.
By the Krein-Friedel formula
the  contribution of scattering states with ${\cal E}>mc^2$
to the integrated density of states is then
\begin{equation}
N_\alpha({\cal E})-N_o({\cal E})= -\frac{1}{2}
\eta(1-\eta)
-\frac{1}{\pi}\arctan\frac{\sin(\eta\pi)}{\cos(\eta\pi)+(-1)^n\tan\mu}\cdot
%\frac{1}{\pi}\left\{
%\begin{array}{cc}
%\delta^d &\hspace*{0.8cm} \eta<1/2\\
%\delta^u &\hspace*{0.8cm}\eta>1/2
%\end{array}
%\right.
\label{dintsing}
\end{equation}
Because of (\ref{tan}) there is again a
typical resonance in the relativistic case for $0<\eta<1/2$ which,
however, for $1/2<\eta<1$ disappear.
When $\eta=1/2$ the resonance is shifted to
infinity. For negative magnetic field the resonance appears
for $1/2<\eta<1$ and disappears for $0<\eta<1/2$.

For $-{\cal E}<-mc^2$ the scattering states are given
by $\Psi_{-{\cal E};l}(t,r,\phi)=\Psi_{{\cal E};l}^*(t,r,\phi)|_{m\rightarrow
-m}$ supplemented with $\tan\mu\rightarrow-\tan\mu$.
The former relation is a consequence of $h_m^*|_{m\rightarrow -m}=-h_m$ for
the radial part
$h_m$ of the Dirac operator and can be verified also by direct calculations.
The latter ensures the same boundary conditions at the origin as for
the energy ${\cal E}$ scattering states, i.e.,
the both sets are in the same Hilbert space. Formally, the bound states with
the energy $\pm E$ and the scattering states with
 $\pm\mu$ are eigenstates of $h_m$, but more
precise look shows that they belong to different self-adjoint
extensions of $h_m$.

In contrast to the Schr\"{o}dinger equation it is
%difficult to identify the physics which underlines self-adjoint
%extensions with the bound state. It is
impossible to find a bound
state with the homogeneous field regularization and our argument
with the decoupling of bound states does not work.
Whenever the magnetic moment coupling induces an {\em attractive}
$V(r)=-g_m\alpha/R^2$ potential inside the flux tube
for the {\em down} component then
it induces a {\em repulsive} $-V$ potential for the {\em up} component.
But it is known that for the bound state to exist it has to be at least
the attractive potential (\ref{tubpot}). An arbitrary weak attractive
potential {\em cannot} lead to bound states (cf. \cite{Hag}).
The same situation persists with the cylindrical shell regularization.
Nevertheless, in the presence of
the threshold state the total energy of the system with filled Dirac sea
changes by infinite positive amount which reflects the stability
of the system.

Note that (\ref{dirham}) is nothing but radial part
of the Klein-Gordon Hamiltonian with the Pauli coupling
\cite{St} and the  relevant results for it are a simple consequence
of the above calculations.

{\bf 4.} {\em Discussion, three dimensional case}.-
We have derived the quantum-mechanical  criterion of instability
(\ref{instab})
which only involves fundamental parameters of matter
and, in particular, shows an instability of massless charged particles
(cf.  \cite{VG} which claims
they cannot exist in nature as they are completely
locally screened in the process of formation).
We did not find any instability of minimally-coupled
relativistic matter without the Abelian Chern-Simons term
(cf. \cite{YH} which
claims that in its presence magnetic field is
spontaneously generated).

In 3D the function $X(\alpha,g_m)$ is replaced
by $X(\alpha,g_m)/L$, with $L$ the length of the flux string.
Therefore the above instability may survive up to 3D
provided the density of states for a long flux ring
preserves essential features of the AB potential.
However, in the case of the 3D Dirac equation
any question about the instability
due to the symmetry of its spectrum is pointless
(cf. the suggestion of \cite{GLP} for `flux spaghetti' vacuum in the
spirit of \cite{NO} as a mechanism for avoiding the
divergence of perturbative QED).
With the AB potential  the pairs of components
$(\chi^1,\chi^4)$, and $(\chi^3,\chi^2)$ of the four-spinor
in the standard representation combine to the components of two bispinors
which satisfy 2D Dirac equations with the opposite sign of mass.
Because $h^*_m=-h_{-m}$ for the radial
part of the 2D Dirac Hamiltonians (cf. (3) of \cite{PG}),
the symmetry of the spectrum (including the threshold and bound states)
of the 3D Dirac equation with respect
to the origin is restored, in accord with the general result \cite{Th}.
A more general discussion of this and related problems will be given
elsewhere \cite{CMO}.

I should like to thank A. Comtet, Y. Georgelin, S. Ouvry, and J. Stern
for many useful and stimulating discussions.
%%%%%%%%%%%%%%%%%%%%%%%%%%%%%%%%%%%%%%%%%%%%%%%%%%%%%%%%%%%%%%%%%%%%%%%%
%\np


\begin{thebibliography}{99}
\bibitem{HSS}
We refer to the diamagnetic inequality when the energy of field and matter
are considered altogether. Sometimes the notion of the diamagnetic and
the paramagnetic inequality is only used for the energy of matter.
See  E. Seiler, {\em Gauge Theories as a Problem of
Contructive Quantum Field Theory and Statistical Mechanics}, LNP 159,
(Springer, Heidelberg, 1982) p. 22 and references therein.
\bibitem{OO}C. R. Hagen, Phys. Rev. Lett.\ {\bf 64}, 2347 (1990);
V. V. Semenov, J. Phys. A: Math. Gen.\ {\bf 25}, L619-L621 (1992);
O. Ogurisu, Lett. Math. Phys. {\bf 30}, 173 (1994).
\bibitem{GJ}P. de Sousa Gerbert and R. Jackiw, Commun. Math. Phys.\ {\bf 124},
 229 (1989).
\bibitem{AW}M. Alford and F. Wilczek, Phys. Rev. Lett.\ {\bf 62}, 1071 (1989).
\bibitem{AB}
W. Ehrenberg and R. E. Siday, Proc. Phys. Soc. {\bf 62B}, 8 (1949);
Y. Aharonov and D. Bohm, Phys. Rev.\ {\bf 115}, 485 (1959);
W. C. Henneberger, Phys. Rev. \ {\bf A 22}, 1383 (1980).
\bibitem{R}S. N. M. Ruijsenaars, Ann. Phys.\ {\bf 146}, 1 (1983).
\bibitem{AS}M. Abramowitch and I. A. Stegun, {\em Handbook of Mathematical
Functions} (Dover Publ., 1973).
\bibitem{RR}R. D. Richtmayer, {\em Principles of Advanced Mathematical Physics}
(Springer, New York, 1978), vol. 1, ch. 10.15.
\bibitem{MRS}E. C. Marino, B. Schroer, and J. A. Swieca, Nucl. Phys.\
{\bf B200}[FS4],
473 (1982). There are some misprints and weak points in their calculations
but, nevertheless, the cited result is valid.
\bibitem{GR}I. S. Gradshteyn and I. M. Ryzhik, {\em Table of Integrals, Series,
and Products} (Academic Press, New York, 1966).
\bibitem{CGO}A. Comtet, Y. Georgelin, and S. Ouvry, J. Phys. A: Math. Gen.\
{\bf 22}, 3917 (1989).
\bibitem{F}J. S. Faulkner, J. Phys. C: Solid State Phys.\ {\bf 10}, 4661
(1977).
\bibitem{PG}P. de Sousa Gerbert, Phys. Rev. \ {\bf D40}, 1346 (1989).
Since in the $l=-(n+1)$-th channel the respective order
of the Bessel functions for the up/down component is 1 and 0
when $\eta=0$ we have taken $\nu=1-\eta$ instead of
$\nu=\eta-1$.
\bibitem{BV}M. Bordag and S. Voropaev, J. Phys. A: Math. Gen.\ {\bf 26},
7637 (1993).
We use $e=-|e|$ which leads to $l\rightarrow -l$ when compared to their
notation.
\bibitem{AC}Y. Aharonov and A. Casher, Phys. Rev. \ {\bf A19}, 2461 (1979).
\bibitem{BV1}We have found that some conclusions of \cite{BV} are not valid:
their coefficients $\alpha_i$ might be negative when they claim them to be
positive; there is an error in calculating the number of bound states.
\bibitem{Hag}C. R. Hagen, Phys. Rev. Lett.\ {\bf 64}, 503 (1990).
%\bibitem{CM}in preparation.
\bibitem{Th}B. Thaler, {\em The Dirac equation} (Springer, New York, 1992) ch.
5. Unfortunately, he does not discuss the bound states.
\bibitem{St}I. I. Kogan, Phys. Lett.\ B{\bf 262}, 83 (1991);
 J. Stern, Phys. Lett.\ B{\bf 265}, 119 (1991).
\bibitem{VG}V. N. Gribov, Nucl. Phys.\ {\bf B206}, 103 (1982).
\bibitem{YH}Y. Hosotani, Phys. Lett.\ B{\bf 319}, 332-318 (1993);
UMN-TH-1238/94 (hep-th/9402096).
\bibitem{GLP}A. S. Goldhaber, H. Li, and R. R. Parwani, ITP-SB-92-40
(hep-th/9305007).
\bibitem{NO}H. B. Nielsen and P. Olesen, Nucl. Phys.\ {\bf B160}, 380 (1979).
\bibitem{CMO}A. Comtet, A. Moroz, and S. Ouvry, IPNO/TH 94-30.
%\input abpro
\end{thebibliography}
\end{document}